\documentclass[aps,prx,reprint,amsmath,amssymb,superscriptaddress,twocolumn]{revtex4}
\usepackage{graphicx}
\usepackage{xcolor}
\usepackage{subfigure}
\usepackage{hyperref,hypcap}
\usepackage{braket}
\usepackage{amsmath}

\begin{document}
\title{Higgs-like modes in two-dimensional spatially-indirect exciton condensates}
\author{Fei Xue}
\affiliation{Department of Physics, University of Texas at Austin, Austin TX 78712, USA} 
\affiliation{Institute for Research in Electronics and Applied Physics \& Maryland Nanocenter,	University of Maryland, College Park, MD 20742, USA}
\author{Fengcheng Wu}
\affiliation{Department of Physics, University of Texas at Austin, Austin TX 78712, USA} 
\affiliation{Condensed Matter Theory Center and Joint Quantum Institute, Department of Physics, University of Maryland, College Park, Maryland 20742, USA} 
\author{A.H. MacDonald}
\affiliation{Department of Physics, University of Texas at Austin, Austin TX 78712, USA} 
\date{\today}

\begin{abstract}
Higgs-like modes in condensed-matter physics have drawn attention 
because of analogies to the Higgs bosons  
of particle physics. Here we use a microscopic time-dependent mean-field theory to 
study the collective mode spectra of two-dimensional spatially indirect exciton 
(electron-hole pair) condensates, focusing on the Higgs-like modes, i.e., those that have a large weight in electron-hole pair amplitude response functions.
We find that in the low exciton density (Bose-Einstein condensate) limit, the dominant Higgs-like modes of spatially indirect exciton condensates 
correspond to adding electron-hole pairs that are orthogonal to the condensed pair state.
We comment on the previously studied Higgs-like collective excitations of superconductors
in light of this finding.  
\end{abstract}

\maketitle

\section{Introduction}
The standard model of particle physics posits a bosonic Higgs field that provides elementary particles 
with mass by breaking symmetries that would otherwise be present.  The recent experimental 
detection \cite{ATLAS2012,CMS2012} of Higgs particles, the elementary excitations of the Higgs field, 
is therefore an important advance in fundamental physics.  
Partly because of their importance to the foundations of physics writ large, there has also been interest 
in excitations that are analogous to Higgs particles in condensed matter, especially in superconducting metals \cite{Varma2015}. Indeed, the absence of massless Goldstone boson excitations in superconductors\cite{Anderson1958,Anderson1963} 
in spite of their broken gauge symmetry, played an important role historically
in the theoretical work \cite{Englert1964,Higgs1964,Kibble1964} that led to the Higgs field proposal.

Emergent symmetry-breaking bosonic fields are common in condensed matter,
where they typically arise from interactions among underlying fermionic fields.
Both electron-electron pair fields, which condense in superconductors, and 
electron-hole pair fields, which condense in ferromagnets and 
in spin or charge density wave systems, are common.  
An important difference between the Higgs fields of particle physics 
and the symmetry-breaking bosonic fields in condensed matter is 
the absence, in the former case, of an understanding of the field's origin 
in terms of underlying degrees of freedom that might be hidden at present, akin to the 
understanding in condensed matter that the order parameter field in a superconductor measures electron-electron pair amplitudes.
Such an understanding might eventually be achieved, 
and analogies to the observed properties of condensed matter 
might once again be valuable in suggesting theoretical possibilities. 
Motivated partly by that hope and partly by the goal of 
shedding new light on the interesting literature \cite{Klein1980,Varma1982,Arovas2011,Barlas2013,Shimano2013,Shimano2014,Sakuto2014,Volovik2014,sherman2015higgs}
on Higgs-like excitation in superconductors and in other condensed-matter 
systems \cite{BEC-BCS1997,BEC-BCS1999,Boehm2008,BEC-BCS2009,endres2012higgs,merchant2014quantum,BEC-BCS2015,Lu2016,jain2017higgs,Esslinger2017,Sun2020}, we address the Higgs-like excitations of two-dimensional spatially indirect 
exciton condensates.

\begin{figure}[htbp]
	\includegraphics[width=1.0\columnwidth]{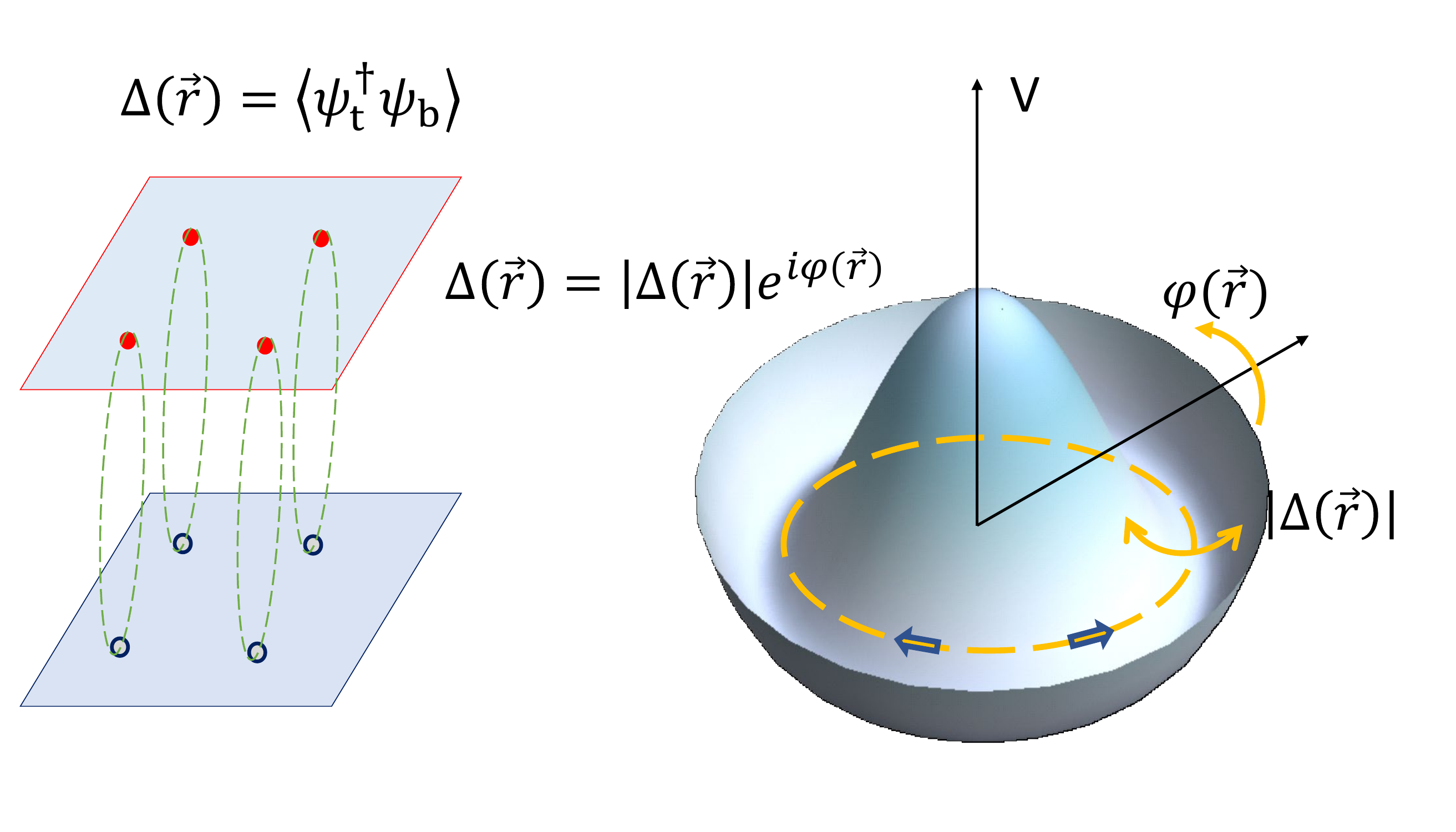}
	\caption{ Schematic illustration of bilayer exciton condensates and of Higgs-like amplitude mode 
		excitations with a Mexican-hat potential.
	}
	\label{Fig:schematic}
\end{figure}

Spatially indirect exciton condensates (SIXCs) are equilibrium or quasiequilibrium states of matter that 
have been extensively studied over the past couple of decades in semiconductor bilayer quantum wells \cite{Keldysh1965,Lozovik1976,Comte1982,Zhu1995},
including in the quantum Hall regime \cite{Eisenstein2004,nandi2012exciton}.
SIXCs have recently been observed 
in van der Waals heterojunction two-dimensional bilayer materials 
both in the presence \cite{Kim2017,Dean2017}
and in the absence \cite{Tutuc2018,Wang2019} of external magnetic fields.  
The bosonic order parameter field 
of a spatially indirect exciton condensate 
\begin{equation}
\label{eq:BosonField}
\Delta(\vec{r}) =  \Psi_{\rm{t}}^{\dagger}(\vec{r}) \Psi_{\rm{b}}(\vec{r}) 
\end{equation} 
has a nonzero expectation value in the broken-symmetry ground state, which 
is characterized by spontaneous interlayer phase coherence
and a suite of related anomalous transport properties \cite{Su2008}.
The labels ${\rm b,t}$ on the 
field operators in 
Eq.~\ref{eq:BosonField} refer to electrons in the bottom ($\rm{b}$) and top ($\rm{t}$) layers of a bilayer 
two-dimensional electron system, as illustrated schematically in Fig.~\ref{Fig:schematic}.  
The SIXC state can be described approximately using a 
mean-field theory \cite{Keldysh1965,Lozovik1976,Comte1982,Zhu1995,Wu&Xue2015} analogous to the 
Bardeen-Cooper-Schrieffer mean-field theory \cite{BCS} of superconductors.  

Superconductors break an exact gauge symmetry related to conservation of the electron number in the many-body
Hamiltonian.  In electron-hole pair condensates the corresponding symmetry is only approximate but 
becomes accurate when the electrons and holes are selected from two different subsets of the single-particle 
Hilbert space whose electron numbers are approximately conserved separately.  
In the case of spatially indirect exciton condensates, the electrons and holes are selected from 
separate two-dimensional layers.  Exceptionally among electron-hole pair condensates, the 
Hamiltonian terms that break separate particle-number conservation can be made 
arbitrarily weak simply by placing an insulating barrier between the two-dimensional subsystems. 
Phenomena associated with broken gauge symmetries can be 
realized as fully as desired by suppressing single-particle processes that allow electrons to move 
between $\rm{b}$ and $\rm{t}$ layers.  The properties of spatially indirect exciton condensates 
are therefore very closely analogous to those of two-dimensional superconductors, as we shall emphasize 
again below.  The main difference between the two cases is that the condensed pairs are charged in
the superconducting case, altering how the ordered states interact with electromagnetic fields.  

In this paper we employ a time-dependent mean-field weak-coupling theory description of
the bilayer exciton condensate's elementary excitations to identify Higgs-like modes and to
demonstrate that in the low-density Bose-Einstein condensate (BEC) limit they have a simple interpretation
as excitations in which electron-hole pairs are added to the system 
in electron-hole pair states that are orthogonal to the $1s$ pair state that is condensed in the many-body
ground state.
In Sec.~\ref{sec:theory} we first briefly describe some details of our theory 
of the SIXC's collective excitations.  In Sec.~\ref{sec:results}
we summarize and discuss numerical results we have obtained
by applying this theory to bilayer two-dimensional electron-hole systems.
Finally, in Sec.~\ref{sec:discussion} we conclude by commenting on  
similarities and differences between bilayer exciton condensates and other systems 
in which Higgs-like modes have been proposed and observed.

\section{Collective Excitation Theory}
\label{sec:theory}
The mean-field theory of the bilayer exciton condensate is 
a generalized Hartree-Fock theory in which translational symmetry
is retained but spontaneous interlayer phase coherence, which breaks separate conservation of the particle number in the two layers, is allowed.  
In Ref.~[\onlinecite{Wu&Xue2015}] we presented a theory of the bilayer exciton condensate's elementary collective 
excitations and quantum fluctuations that accounts for quadratic variations 
of the Hartree-Fock energy functional.  Importantly for the findings that are the focus 
of the present paper, the theory fully accounts 
for the long-range Coulomb interactions among electrons and holes.
Theories that do not recognize the Coulomb's interactions' long range or 
which do not treat electrostatic and exchange interactions on an equal footing
make qualitative errors in describing spatially indirect exciton condensates.
This comment applies in particular to the short-range interaction models that can be conveniently analyzed 
using Hubbard-Stratonovich transformations (see, for example, Ref.~\cite[pp.~333--335]{Negele1988}).
In this section we briefly summarize that theory and generalize it in a way that makes 
evaluation of the two-particle Green's 
functions that characterize the system's particle-hole excitations particularly convenient.  
The SIXC's elementary excitation energies are identified with the
poles of those Green's functions and are the eigenvalues of a matrix constructed 
from the kernel of the quadratic-fluctuation energy functional.  The character
of given elementary excitations is classified by determining which particle-hole pair response 
functions have large residues at its poles.

\subsection{Mean-field theory} 
 
For simplicity we neglect the spin and valley degrees of 
freedom that often play a role in realistic SIXC systems.  
The self-consistent Hartree-Fock 
mean-field Hamiltonian of the broken-symmetry bilayer exciton condensate state is then \cite{Wu&Xue2015}
\begin{equation}
H_{MF}=\sum_{\vec{k}}(a_{c\vec{k}}^{\dagger}, a_{v\vec{k}}^\dagger)
(\zeta_{\vec{k}}+\xi_{\vec{k}} \sigma_z-\Delta_{\vec{k}} \sigma_x)
\begin{pmatrix} a_{c\vec{k}}\\a_{v\vec{k}}\end{pmatrix}.
\label{eq:MF}
\end{equation}
Here $a_{n\vec{k}}$ and $a^{\dagger}_{n\vec{k}}$ are fermionic annihilation and creation operators for the 
conduction ($n=c$) band electrons localized in the top layer 
and valence ($n=v$) electrons localized in the bottom layer, $\sigma_{z,x}$ are Pauli matrices that act in the band space, and
$\zeta_{\vec{k}}=\hbar^2k^2[1/(4m_e)-1/(4m_h)]$ accounts for the difference between
conduction and valence band effective masses, which plays no role in the temperature 
$T=0$, charge-neutral limit that we consider.  
For convenience, we take $m_e=m_h $ in the calculations described below.
The dressed band parameters $\xi_{\vec{k}}$ and $\Delta_{\vec{k}}$ are 
obtained by solving the self-consistent-field equations:
\begin{equation}
\label{eq:SC}
\begin{split}
&\xi_{\vec{k}}=\frac{\hbar^2k^2}{4m} 
+ \frac{\tilde{\mu}}{2}-\frac{1}{2A}\sum_{\vec{k}'}V_{\vec{k}-\vec{k}'}(1-\xi_{\vec{k}'}/E_{\vec{k}'}),\\
&\Delta_{\vec{k}}=\frac{1}{2A}\sum_{\vec{k}'}U_{\vec{k}-\vec{k}'}\frac{\Delta_{\vec{k}'}}{E_{\vec{k}'}},\\
&E_{\vec{k}}=\sqrt{\xi_{\vec{k}}^2+\Delta_{\vec{k}}^2},
\end{split}
\end{equation}
where $m=m_em_h/(m_e+m_h)$ is the reduced mass and $A$ is the area of the two-dimensional
system.  In Eq.~\ref{eq:SC}, $V_{\vec{q}}=2\pi e^2/(\epsilon q)$ and $U_{\vec{q}}=V_{\vec{q}} \, \exp(-qd)$ 
are the intralayer and interlayer Coulomb interactions,
\begin{equation}
\begin{split}
&\tilde{\mu}=\mu+4\pi e^2 n_{ex} d /\epsilon,\\  
&n_{ex}=\frac{1}{2A}\sum_{\vec{k}}(1-\xi_{\vec{k}}/E_{\vec{k}}),
\end{split}
\label{eq:SC_2}
\end{equation}
$\mu$ is the chemical potential parameter for excitons, and $n_{ex}$ is equal to both the density of 
conduction band electrons and the density of valence band holes.   Below we refer to 
$n_{ex}$ as the density of excitons;  this terminology 
is motivated mainly by the low-density limit in which $n_{ex} a_B^{*2} \ll 1$.
(Here $a_B^*= \hbar^2\epsilon/me^2$ is the Bohr radius, which is the bound electron-hole pair size in the limit of small layer separations.)
The exciton chemical potential parameter $\mu=E_g - V_b$ can be adjusted electrically by applying a gate voltage to alter 
the spatially indirect band gap $E_g$, provided that the barrier between conduction and 
valence band layers is sufficiently opaque,
or by applying a bias voltage $V_b$ between layers \cite{Wang2019}.

The mean-field ground state is 
\begin{equation}
\label{eq:groundstate}
\ket{XC}=\prod_{\vec{k}}\gamma^{\dagger}_{\vec{k},0}\ket{0}=
\prod_{\vec{k}}(u_{\vec{k}}a_{c\vec{k}}^{\dagger}+v_{\vec{k}}a_{v\vec{k}}^{\dagger})\ket{0},
\end{equation}
where  
\begin{equation}
\label{eq:ukvk}
	u_{\vec{k}}=\sqrt{\frac{1}{2}(1-\xi_{\vec{k}}/E_{\vec{k}})}, \quad
	v_{\vec{k}}=\sqrt{\frac{1}{2}(1+\xi_{\vec{k}}/E_{\vec{k}})},
\end{equation}
and $\gamma_{\vec{k},0}^{\dagger}$ is the creation operator for the dressed valence 
band quasiparticle states that are occupied in $\Ket{XC}$.
Note that we have chosen $u_{\vec{k}}$ and $v_{\vec{k}}$ to be real and that there 
is a family of degenerate states that differ only by a global shift in the 
phase difference between electrons localized in different layers.

\subsection{Quadratic fluctuations}
We construct our theory of quantum fluctuations and collective excitations
by starting from a many-body state that incorporates arbitrary single-particle-hole excitation
corrections to the mean-field state:
\begin{equation} 
\label{eq:phi}
\Ket{\Phi}= \prod_{\vec{k}} \Big[\mathcal{Z}_{\vec{k}} +\sum_{\vec{Q}}z_{\vec{k}}(\vec{Q})
\gamma_{\vec{k}+\vec{Q},1}^{\dagger}\gamma_{\vec{k},0}\Big]\Ket{XC},
\end{equation} 
where $\gamma_{\vec{k},1}^{\dagger}$ is a creation operator
for a quasiparticle state in the band that is empty in $\Ket{XC}$:
\begin{equation}
\gamma_{\vec{k},1}^{\dagger}=v_{\vec{k}}a_{c\vec{k}}^{\dagger}-u_{\vec{k}}a_{v\vec{k}}^{\dagger}
\end{equation}
and
\begin{equation}
\mathcal{Z}_{\vec{k}} =\sqrt{1-\sum_{\vec{Q}}|z_{\vec{k}}(\vec{Q})|^2}
\end{equation}
is a normalization factor.  The complex  parameters $z_{\vec{k}}(\vec{Q})$
are the amplitudes of all possible single-particle-hole excitations.

To characterize the quantum fluctuations of the mean-field state in a
physically transparent way, we define the observables
\begin{equation}
\label{eq:tau}
\hat{\tau}_{\alpha=\{x,y,z\}}(\vec{Q})=\frac{1}{2}\sum_{\vec{k}}(a_{c\vec{k}+\vec{Q}}^{\dagger}, a_{v\vec{k}+\vec{Q}}^\dagger)
\, \sigma_{\alpha}
\begin{pmatrix} a_{c\vec{k}}\\a_{v\vec{k}}\end{pmatrix}.
\end{equation}
Note that $\langle \Phi | \tau_{\alpha}(\vec{Q}) | \Phi \rangle = 
\langle \Phi | \tau_{\alpha}(-\vec{Q}) | \Phi \rangle^*$. 
For the interlayer phase choice we have made, the 
mean-field value of the order parameter $\Delta^{MF}$ is real and spatially constant:
\begin{equation}
\Delta^{MF} = \frac{1}{A} \sum_{\vec{k}} u_{\vec{k}} v_{\vec{k}},
\end{equation}
where $A$ is the sample area.  When fluctuations are included, the order parameter becomes
\begin{widetext}
\begin{equation} 
\Delta(\vec{r}) =  
 \frac{1}{A} \sum_{\vec{Q}}  |\langle \tau_x(\vec{Q}) \rangle | \cos(\vec{Q}\cdot\vec{r} - \varphi_{\vec{Q}x})   + i 
|\langle \tau_y(\vec{Q}) \rangle |  \cos(\vec{Q}\cdot\vec{r} - \varphi_{\vec{Q}y}) 
\end{equation}
\end{widetext}
where $\langle \ldots \rangle = \langle \Phi | \ldots | \Phi \rangle$ and  
$\varphi_{\vec{Q}\alpha}$ is defined by $\langle \tau_{\alpha}(\vec{Q}) \rangle = 
|\langle \tau_{\alpha}(\vec{Q}) \rangle | \exp(i\varphi_{\vec{Q}\alpha})$.
It follows that, to leading order, fluctuations in the order parameter magnitude are proportional to
$\langle \tau_{x}(\vec{Q})\rangle $, while fluctuations in the order parameter phase are related to 
$\langle \tau_{y}(\vec{Q})\rangle$.  $\langle \tau_{z}(\vec{Q}) \rangle$ measures fluctuations
in the exciton density.  

We quantize fluctuations in the XC state by constructing the Lagrangian:
\begin{equation}
\mathcal{L}=\langle \Phi | i\hbar \partial_t -H|\Phi \rangle \approx \mathcal{B}-\delta E^{(2)},
\label{eq:lag}
\end{equation}
where $\delta E^{(2)}$ is the harmonic fluctuation energy 
functional \cite{Giuliani2005} 
and $\mathcal{B} = \langle \Phi | i\hbar \partial_t |\Phi \rangle$ is the Berry phase term which 
enforces bosonic quantization rules on the 
$z_{\vec{k}}(\vec{Q})$ fluctuation parameters.  
The energy functional is obtained by taking the expectation 
value of the many-body Hamiltonian and has the form
\begin{widetext}
	\begin{equation}
	\delta E^{(2)}= \langle \Phi |H|\Phi \rangle = \sum_{\vec{Q},\vec{k},\vec{p}}\{\mathcal{E}_{\vec{k},\vec{p}}(\vec{Q})z_{\vec{k}}^*(\vec{Q})z_{\vec{p}}(\vec{Q})
	+\frac{1}{2}\Gamma_{\vec{k},\vec{p}}(\vec{Q})
	[z_{\vec{k}}(\vec{Q})z_{\vec{p}}(-\vec{Q})+z^*_{\vec{k}}(\vec{Q})z^*_{\vec{p}}(-\vec{Q})]\}.
	\label{eq:EF}
	\end{equation}
\end{widetext}
$\delta E^{(2)}$ accounts for variations in band kinetic energy and 
Hartree and exchange interaction energy as 
the many-electron state fluctuates.  Explicit forms for the 
matrices $\mathcal{E}$ and $\Gamma$ are given in Appendix ~\ref{App1}. 
 
We separate the fluctuation Hamiltonian into amplitude and phase 
fluctuation contributions by making the 
change in variables
\begin{align}
z_{\vec{k}}(\vec{Q})&=\frac{1}{\sqrt{2}}[x_{\vec{k}}(\vec{Q})+\mathrm{i}y_{\vec{k}}(\vec{Q})],\\ 
z^*_{-\vec{k}}(-\vec{Q})&=\frac{1}{\sqrt{2}}[x_{\vec{k}}(\vec{Q})-\mathrm{i}y_{\vec{k}}(\vec{Q})].
\end{align} 
Note that $x_{\vec{k}}(\vec{Q})$ and $y_{\vec{k}}(\vec{Q})$ are also complex, but satisfy
$x_{\vec{k}}(\vec{Q}) = x^*_{-\vec{k}}(-\vec{Q})$ and $y_{\vec{k}}(\vec{Q}) = y^*_{-\vec{k}}(-\vec{Q})$ so
that $\vec{Q}$ and $-\vec{Q}$ fluctuations are not independent.
Order parameter amplitude and exciton density 
fluctuations are both related to fluctuations in the $x$ fields, while 
phase fluctuations are related to fluctuations in the $y$ fields:
\begin{widetext}
\begin{equation}
\label{eq:tau2}
\begin{aligned}
&\tau_{x}(\vec{Q})=\langle \Phi | \hat{\tau}_x((\vec{Q})|\Phi \rangle=\frac{1}{\sqrt{2}}\sum_{\vec{k}}(v_{\vec{k}}v_{\vec{k}+\vec{Q}}
	-u_{\vec{k}}u_{\vec{k}+\vec{Q}}) \, x_{\vec{k}}(\vec{Q}),\\
&\tau_{y}(\vec{Q})=\langle \Phi | \hat{\tau}_y(\vec{Q})|\Phi \rangle=\frac{1}{\sqrt{2}}\sum_{\vec{k}}(v_{\vec{k}}v_{\vec{k}+\vec{Q}}
	+u_{\vec{k}}u_{\vec{k}+\vec{Q}})\, y_{\vec{k}}(\vec{Q}),\\	
&\tau_{z}(\vec{Q})=\langle \Phi | \hat{\tau}_z(\vec{Q})|\Phi \rangle=\frac{1}{\sqrt{2}}\sum_{\vec{k}}(u_{\vec{k}}v_{\vec{k}+\vec{Q}}
	+v_{\vec{k}}u_{\vec{k}+\vec{Q}})\, x_{\vec{k}}(\vec{Q}).
\end{aligned}
\end{equation}
\end{widetext}
Note that although $\tau_{x}(\vec{Q})$ and exciton 
density $\tau_{z}(\vec{Q})$ fluctuations are both related to $x_{\vec{k}}(\vec{Q})$,
they have different $\vec{k}$-dependent weighting factors.  
For each wave vector transfer $\vec{Q}$ we define vectors of $x$ and $y$ variables,  
$\mathbf{X}(\vec{Q}) \equiv (x_{\vec{k}_1}(\vec{Q}),\ldots,x_{\vec{k}_i}(\vec{Q}),\ldots)$ and 
$\mathbf{Y}(\vec{Q}) \equiv (y_{\vec{k}_1}(\vec{Q}),\ldots,y_{\vec{k}_i}(\vec{Q}),\ldots)$, 
whose elements are labeled by the particle-hole pair's hole momentum.  In terms of these vector 
variables the action 
\begin{widetext}
\begin{equation}
S=\int dt (\mathcal{B}-\delta E^{(2)})
=\frac{1}{2}\sum_{\vec{Q}}\int dt\Big(\hbar\mathbf{Y}^{\dagger}(\vec{Q}) \partial_t \mathbf{X}(\vec{Q})-\hbar\mathbf{X}^{\dagger}(\vec{Q}) \partial_t \mathbf{Y}(\vec{Q})-\mathbf{X}^{\dagger}(\vec{Q}) \mathcal{K}^{(+)}\mathbf{X}(\vec{Q}) -
\mathbf{Y}^{\dagger}(\vec{Q})\mathcal{K}^{(-)}\mathbf{Y}(\vec{Q}) \Big),
\label{eq:quadS}
\end{equation}
\end{widetext}
where $\mathcal{K}^{(\pm)}_{\vec{k},\vec{p}}(\vec{Q})=
\big(\mathcal{E}_{\vec{k},\vec{p}}\pm \Gamma_{\vec{k},-\vec{p}}\big)(\vec{Q})$
is real and symmetric for both sign choices and 
we have used the fact that $\int dt\mathbf{Y}^{\dagger}\partial_t \mathbf{X}=-\int dt\mathbf{X}\partial_t \mathbf{Y}^{\dagger}$.

Minimizing the action yields the following equations of motion: 
\begin{eqnarray} 
\hbar{\partial_t \mathbf{X}} &=& \mathcal{K}^{(-)} \, \mathbf{Y}(\vec{Q}), \nonumber \\
\hbar{\partial_t \mathbf{Y}} &=& -  \mathcal{K}^{(+)} \,  \mathbf{X}(\vec{Q}).
\end{eqnarray} 
This theory of fluctuations is equivalent to time-dependent Hartree-Fock theory for the 
exciton condensate state response functions.  It is in the same spirit as auxiliary field 
functional integral theories of harmonic quantum fluctuations but unlike those approaches 
treats Hartree and exchange energy contributions on an equal footing \cite{Negele1988},
an attribute that is necessary if the bilayer exciton condensate is to be described directly.  

\subsection{Particle-hole correlation functions} 
Collective modes give rise to poles in particle-hole channel Greens functions.
As in the case of BCS superconductors \cite{Varma1982,Varma2015}, those 
collective modes that have large residues in ($\hat{\tau}_x,\hat{\tau}_x$) particle-hole Greens functions
can be identified as Higgs-like modes.  
Because the fluctuation Hamiltonian $\delta E^{(2)}$ is the sum of quadratic contributions in
the $X$ and $Y$ fields, which are canonically conjugate,
we can apply the generalized Bogoliubov transformation \cite{Xiao2009} described in 
detail below to write the fluctuation Hamiltonian in a free-boson form:
\begin{equation} 
 H=E_0+\sum_{\vec{Q}}\sum_i  \hbar \omega_{i}(\vec{Q}) B_i^{\dagger}(\vec{Q}) B_i(\vec{Q}).
\end{equation} 
where $\hbar \omega_{i}(\vec{Q})$ is an excitation energy and
$B^{\dagger}_i(\vec{Q})$ and $B_i(\vec{Q})$
are linear combinations of the $x_{\vec{k}}$ and $y_{\vec{k}}$ fields.  
To evaluate correlation functions involving the $\tau_{\alpha}(\vec{Q})$ fields 
we reexpress $x_{\vec{k}}(\vec{Q})$ and $y_{\vec{k}}(\vec{Q})$ in Eqs.~\ref{eq:tau2} 
in terms of these free boson fields. The character of collective excitations is revealed by the residues of 
response functions at poles that lie below particle-hole continua. 

To carry out this procedure explicitly, we start from the assumptions that the 
amplitude/density kernel  
$\mathcal{K}^{(+)}$ is positive definite at any $\vec{Q}$ and that the phase kernel $\mathcal{K}^{(-)}$ is positive definite 
for $\vec{Q} \ne 0$ and positive semidefinite for $\vec{Q}=0$.  These assumptions are satisfied whenever
the mean-field condensate is metastable.  The zero eigenvalue of $\mathcal{K}^{(-)}$ at $\vec{Q}=0$ 
arises from the broken $U(1)$ symmetry associated with spontaneous interlayer phase coherence.  
Because $\mathcal{K}^{(+)}$ is real symmetric and positive definite, it is possible \cite{Horn2012}
to perform a Cholesky decomposition for each $\vec{Q}$ by writing 
\begin{equation}
\mathcal{K}^{(+)}=\mathbf{L} \mathbf{L}^{T}
\end{equation}
and then to diagonalize 
\begin{equation}
\label{eq:gamma}
\mathbf{\Gamma}=\mathbf{L}^T \mathcal{K}^{(-)}\mathbf{L}=\mathbf{\Gamma}^T.
\end{equation}
Writing $\mathbf{\Gamma} = \mathbf{S} \mathbf{\Lambda} \mathbf{S}^{T}$, where
$\mathbf{S}$ is an orthogonal matrix and $\mathbf{\Lambda}$ is a diagonal matrix, we define new fields  
\begin{equation}
\begin{aligned}
&\mathbf{\Psi}=\mathbf{S}^T \mathbf{L}^{-1} \mathbf{Y},\\
&\mathbf{\Pi}= \mathbf{S}^T \mathbf{L}^T \mathbf{X},
\end{aligned}
\end{equation}
where $\mathbf{X},\mathbf{Y}$ are the density and phase fluctuation vectors labeled by 
wavevector $\vec{k}$ introduced above.  The transformed Hamiltonian is

\begin{equation}
\begin{aligned}
\mathcal{H}&=\frac{1}{2}(\mathbf{X}^{\dagger}\mathcal{K}^{(+)}\mathbf{X}+\mathbf{Y}^{\dagger}\mathcal{K}^{(-)}\mathbf{Y})\
=\frac{1}{2}(\mathbf{\Pi}^{\dagger}\mathbf{\Pi}+\mathbf{\Psi}^{\dagger}\mathbf{\Lambda}\mathbf{\Psi})\\
&=\frac{1}{2}\sum_i(|\Pi_i|^2+\hbar^2\omega_i^2 |\Psi_i|^2).
\end{aligned}
\end{equation}
Noting that the eigenvalues of $\mathbf{\Gamma}$ are identical to the eigenvalues 
of $\mathcal{K}^{(+)}\mathcal{K}^{(-)}$ and that the equation of motion for phase fluctuations can 
be written in the form 
\begin{equation}
\label{eq:eom}
-\hbar^2\partial_t^2 \mathbf{Y} =  \mathcal{K}^{(+)}\mathcal{K}^{(-)}\mathbf{Y},
\end{equation}
we have identified  them as the 
squares $\omega_i^2$ of the elementary excitation frequencies.

When expressed in terms of the normal mode fields, the action in Eq.~\ref{eq:quadS} has the form
	\begin{equation}
	S=\frac{1}{2}\sum_{\vec{Q}}\int dt\Big(\hbar\mathbf{\Psi}^{\dagger}\partial_t \mathbf{\Pi}-\hbar\mathbf{\Pi}^{\dagger}\partial_t \mathbf{\Psi}
	- \mathbf{\Pi}^{\dagger}\mathbf{\Pi}-\mathbf{\Psi}^{\dagger}\mathbf{\Lambda}\mathbf{\Psi}\Big)_{\vec{Q}}.
	\label{eq:DiagS}
	\end{equation}
The time-ordered Green's function at each $\vec{Q}$ can be calculated directly from the action of fields $\phi$,
\begin{equation}
\label{eq:Greenfunction}
\begin{aligned}
\mathcal{G}(\omega)
&=\begin{pmatrix}
-\mathbf{\Lambda}& i \hbar \omega\\
-i \hbar \omega&-\mathbf{I}
\end{pmatrix}^{-1}\\
&=(\rm{det}|\mathbf{\Lambda}-\hbar^2\omega^2|)^{-1}\begin{pmatrix}
-\mathbf{I}& -i \hbar \omega\\
i \hbar \omega&-\mathbf{\Lambda}
\end{pmatrix}.
\end{aligned}
\end{equation}
Note that the Green's function is a $2\times2$ matrix in the basis of fields $\{\mathbf{\Psi},\mathbf{\Pi}\}$.
The $\tau_{\alpha}$ fields can be expressed in terms of the normal-mode fields for each $\vec Q$ using 
\begin{equation}
\label{eq:tau3}
\begin{aligned}
&\tau_{x}=\sum_i [ \mathbf{T}_{x} (\mathbf{L}^T)^{-1}\mathbf{S}]_i\ \Pi_i \equiv \tau_{x,i} \Pi_i, \\
&\tau_{z}=\sum_i [ \mathbf{T}_{z} (\mathbf{L}^T)^{-1}\mathbf{S}]_i\ \Pi_i \equiv \tau_{z,i} \Pi_i, \\
&\tau_y=\sum_i  (\mathbf{T}_y\mathbf{L}\mathbf{S})_i\, \Psi_i  \equiv \tau_{y,i} \Psi_i,
\end{aligned}
\end{equation}
where the $\mathbf{T}_{\alpha}$ on the right-hand sides of these equations are the matrix forms of Eq.~\ref{eq:tau2}.
The linear response functions are related to the time-ordered Green's functions,
\begin{equation}
\label{eq:LRT_Green}
\chi_{AB}=-\frac{i}{\hbar} \braket{\mathcal{T}[\hat{A}(t),\hat{B}(t')]}.
\end{equation}
The response functions of operators expressed in terms of fields $\mathbf{\Pi_i}$
can be evaluated by performing the average in Eq.~\ref{eq:LRT_Green} 
using the quadratic action weighting factor with the result that
\begin{equation}
\chi_{AB}=\sum_i \frac{\omega_{i}}{2}(\frac{A_{0i}B_{i0}}{\omega-\omega_{i}+\mathrm{i}\eta}-\frac{A_{i0}B_{0i}}{\omega+\omega_{i}+\mathrm{i}\eta}),
\end{equation}
where $A_{mn}$ 
is the matrix element in a complete set of fields $\mathbf{\Pi_i}$. 
We identify Higgs-like modes by finding isolated eigenvalues $|\omega_i|^2$ with large values of $|\tau_{x,i}|^2$ in
the imaginary part of the response functions for positive frequencies:
\begin{equation}
\label{eq:response_1}
\text{Im}\chi_{xx}(\omega) = -\pi \sum_i\frac{\omega_{i}}{2} |\tau_{x,i}|^2  \delta(\omega-\omega_{i}). 
\end{equation}
Note that $\text{Im}\chi_{zz}$ is similar to Eq.~\ref{eq:response_1} by replacing $|\tau_{x,i}|^2$ with $|\tau_{z,i}|^2$. For $\text{Im}\chi_{yy}$, the factor $\omega_i/2$ also needs to be replaced by $1/(2\omega_i)$.

An alternative approach to obtain the same results is to map the diagonalized action in Eq.~\ref{eq:DiagS} 
to that of a set of independent harmonic oscillators, defining the oscillator ladder operators $B_i$ by 
\begin{equation}
\begin{aligned}
&\Pi_i=i \sqrt{\frac{\hbar\omega_i}{2}}(B_i^{\dagger}-B_i)\\
&\Psi_i=\sqrt{\frac{1}{2\hbar\omega_i}}(B_i^{\dagger}+B_i),
\end{aligned}
\end{equation} 
where $B_i$ satisfy
 $[B_i,B_i'^{\dagger}] = \delta_{i,i'}$.
The fluctuation Hamiltonian for each $\vec{Q}$ is then 
\begin{equation}
\label{eq:freeBoson}
\mathcal{H}=E_0+\sum_i \, \hbar \omega_i \, B_i^{\dagger}B_i.
\end{equation}

From the general linear response theory, the Lehmann representation of the response function is \cite{Giuliani2005}
\begin{equation}
\chi_{AB}(\omega)=\frac{1}{\hbar}\sum_{mn}\frac{P_m-P_n}{\omega-\omega_{nm}+i\eta}A_{mn}B_{nm},
\end{equation}
where $P_n=\frac{e^{-\beta E_n}}{\sum_n e^{-\beta E_n}} (\beta=1/k_B T)$ is the occupation probability, 
$\omega_{nm}=(E_n-E_m)/\hbar$ is the excitation frequency, and $A_{mn}\equiv\bra{\psi_m}\hat{A}\ket{\psi_n}$
is the matrix elements in a complete set of exact eigenstates $\ket{\psi_n}$ of $\hat{\mathcal{H}}$.
At zero temperature, the imaginary part of the response function is
\begin{equation}
\label{eq:response}
\begin{aligned}
\text{Im}\chi_{AB}=&-\frac{\pi}{\hbar}\sum_{nm}P_m[A_{mn}B_{nm}\delta(\omega-\omega_{nm})\\
			&-A_{nm}B_{mn}\delta(\omega+\omega_{nm})]\\
			=&-\frac{\pi}{\hbar}\sum_n[A_{0n}B_{n0}\delta(\omega-\omega_{n0})-A_{n0}B_{0n}\delta(\omega+\omega_{n0})].
\end{aligned}
\end{equation}

\section{Results}
\label{sec:results}

We now apply the theory outlined above to bilayer exciton condensates.
The length and energy units we use in our calculations are those appropriate for Coulomb interactions,
the Bohr radius $a_B^*=\epsilon\hbar^2/(m e^2)$, and the effective Rydberg $\rm{Ry}^*=e^2/(2\epsilon a_B^*)$.
Typical values of these parameters in transition metal dichalcogenides\cite{Wu&Xue2015} bilayers 
are $a_B^*\approx10 \text{\AA}, \text{Ry}^*\approx100 \text{meV}$,
while typical values for GaAs bilayer quantum wells \cite{Harrison2005} are $a_B^*\approx100 \text{\AA}, \text{Ry}^*\approx5 \text{meV}$. For all the numerical calculations, we assume $m_e=m_h$, and use $d/a_B^*=0.5$.

The time-dependent mean-field theory is expected to be most reliable in the dilute density limit that resembles a two-dimensional hydrogenlike problem. We consider two cases where the chemical potential parameter $\mu$ is just below the binding energy and $\mu=0$. In both two cases, we use the momentum cutoff $k_c a^*_B=6$ and a $200\times200 k$ mesh. The momentum cutoff is chosen to make sure the exciton density at the cutoff is smaller than $1\times10^{-8}$. Because our momentum space grids are necessarily discrete, corresponding to applying periodic boundary conditions to a
finite area system, the number of particle-hole pairs at a given excitation momentum residing on our k-space grid is finite.
The distinctions between particle-hole continua and isolated collective modes made below 
are qualitative but, for the most part, unambiguous. 

\begin{figure}[htbp]
	\includegraphics[width=1.0\columnwidth]{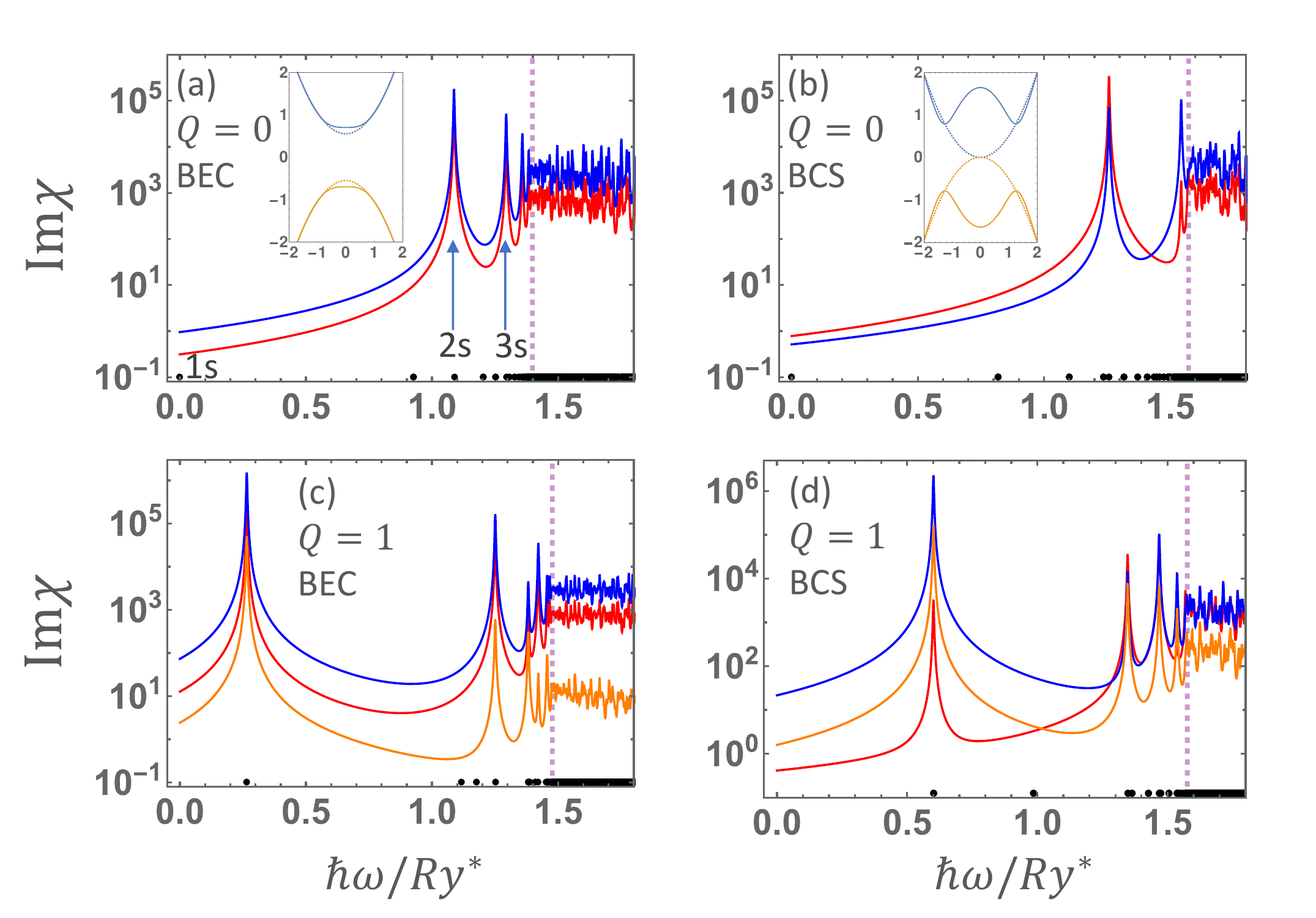}
	\caption{(Color online) 
		Spectra of the magnitude of the imaginary part of $\tau_{x}-\tau_{x}$ (red lines), $\tau_{y}-\tau_{y}$ (blue lines), and $\tau_{z}-\tau_{z}$ (yellow lines) response functions. 
		Black dots along x-axis represent the positive collective excitation energies which are square roots of eigenvalues of $\mathbf{\Gamma}$ [Eq.~\ref{eq:gamma}] and $\mathcal{K}^{(+)}\mathcal{K}^{(-)}$ [Eq.~\ref{eq:eom}].
		The purple dashed lines denote the location of the electron-hole continuum, i.e., the minimum of $E_{\vec{k}}+E_{\vec{k}+\vec{Q}}$. 
		(a) and (b) show the results of $Q a^*_B=0$ at low exciton density ($n_{ex}a_B^{*2}=0.01$) and high exciton density ($n_{ex}a_B^{*2}=0.1$), respectively. Insets in (a) and (b) show the mean-field energy bands (solid lines) and noninteracting bands (dashed lines) as a function of ky (kx is at a fixed value) in these two cases. (c) and (d) show similar results at finite center-of-mass momentum $Q a^*_B=1$. Note that the yellow line for $\text{Im}\chi_{zz}$ is absent at $Q=0$ because its value is zero at all excitation energies.
		}
	\label{Fig:Spectra}
\end{figure}

Bilayer exciton condensates have a BEC-BCS crossover that can be tuned by 
varying not the strength of interactions, \cite{greiner2003emergence,Jin2004,Salomon2004,Zwerger2008,RevModPhys2008,RevModPhys2010} as in cold-atom systems, but
the Fermi energy of the underlying electrons and holes \cite{Perali2013,Liu2017,Dean2017,Neilson2018}.  In two dimensions 
the binding energy of a single electron-hole pair is $ 4 Ry^*$ (d=0), and the Fermi energy in 
$\text{Ry}^*$ units is $ 2 \pi n a_B^{*2}$.  The bilayer exciton 
condensate therefore approaches a BEC limit for small values of $n a_B^2$ when the chemical potential is positive but below the exciton binding energy shown in the inset of Fig.~\ref{Fig:Spectra}(a).  As the chemical potential becomes zero or even negative, the condensate approaches a BCS limit shown in the inset of Fig.~\ref{Fig:Spectra}(b) for $n a_B^2 \gtrsim 0.1$ at $\mu=0$.  
In Figs~\ref{Fig:Spectra}(a) and (c), we plot the magnitude of the imaginary part of response functions $\text{Im}\chi_{xx}$ (red lines), $\text{Im}\chi_{yy}$ (blue lines), and $\text{Im}\chi_{zz}$ (yellow lines) as a function of positive excitation energies for $\vec{Q}=0$ and $\vec{Q}a_B^*=1$ for a low exciton density in the BEC regime, 
$n_{ex}a_B^{*2}=0.01$. Black dots along the $x$ axis represent discrete collective mode spectra $\omega_i$, and we denote the particle-hole continuum (the minimum of $E_{\vec{k}}+E_{\vec{k}+\vec{Q}}$) with a vertical purple dashed line. To get smooth lines, we use Lorentzian functions to plot the Dirac-$\delta$ functions (Eq.~\ref{eq:response_1}) with a width equal to the smallest energy scale in our calculation, i.e., $k_{mesh}^2/2$. These calculations identify certain collective modes at energies below 
the particle-hole continuum  that have a large weight in the $(\tau_x,\tau_x)$ pair amplitude response functions.
This result is reminiscent of the finding in earlier work \cite{Varma1982,Varma2015}
that for superconductors there is a collective mode 
at the edge of the excitation continuum with a large residue in the pair amplitude response
function. At finite excitation wavevector $\vec{Q}$ additional modes have significant 
pair amplitude character.  

\begin{figure*}[htbp]
	\includegraphics[width=1.8\columnwidth]{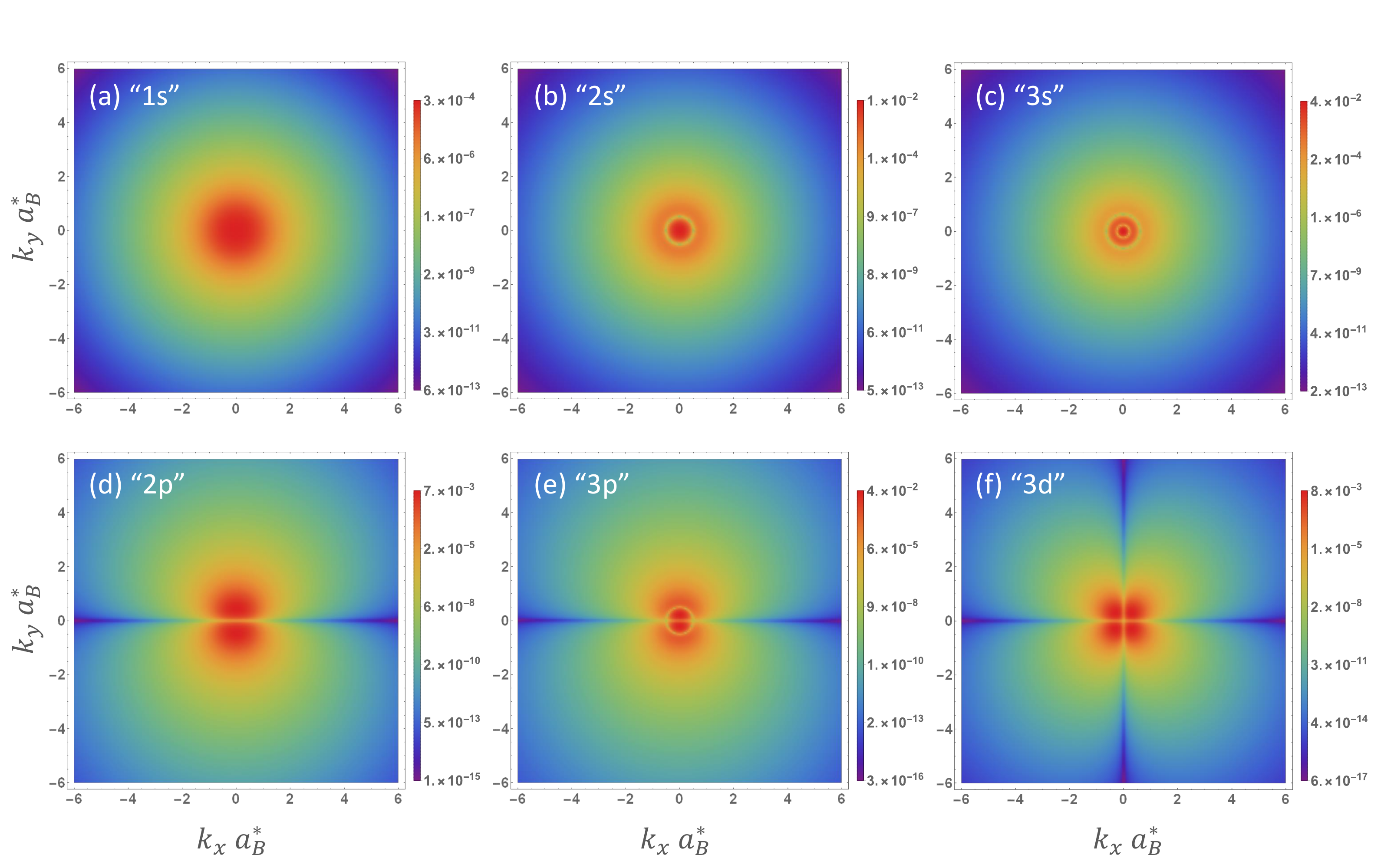}
	\caption{ (Color online) The squared modulus of wavefunctions of collective modes on the momentum grid at $\vec{Q}=0$ in the BEC regime. (a) to (c) show $ns$-like atomic orbital distribution corresponding to gapless Goldstone modes and the two Higgs-like modes denoted with arrows in Fig.~\ref{Fig:Spectra}(a). These three modes all have isotropic momentum distributions but with different numbers of nodes (the number of nodes is equal to $n-1$). (d) to (f) show $2p$, $3p$, and $3d$-like atomic orbitals distributions which do not contribute to the amplitude response functions. The energy sequence of all six collective modes is $1s, 2p, 2s, 3d, 3p$, and $3s$. Note that $2p, 3p$, and $3d$ are doubly degenerate, and one of each doublet is plotted here.
	}
	\label{Fig:BECwfn}
\end{figure*}                                    
The corresponding results for $\chi_{yy}$ (exciton phase) and $\chi_{zz}$ (exciton density) are presented as blue and yellow lines in Fig~\ref{Fig:Spectra}. All three lines share similar peak positions due to the coupling between different channels except that $\chi_{zz}$ is absent in $Q=0$. The differences in coupling between exciton density fluctuations and amplitude fluctuations is due to the different $\vec{k}$-dependent weighting factors in Eq.~\ref{eq:tau2} although both responses are related to the changes in the $x_{\vec{k}}(\vec{Q})$ fields.
The strong mixing of phase and amplitude fluctuations is also observed in superconductors with particle-hole symmetry breaking \cite{BEC-BCS2015}.
The Goldstone mode energy vanishes as $\vec{Q} \to 0$ in the 
electron-hole pair case because these modes are neutral, whereas it has a finite energy in the
three-dimensional electron-electron pair case of superconductors because of the divergence in the 
Coulomb interactions as $\vec{Q} \to 0$.  

The collective modes that have large weight 
in the $\chi_{xx}$ response function are entirely different in character. At $\vec{Q}=0$, the Goldstone mode contribution to any response functions $\text{Im}\chi_{\alpha\alpha}$ vanishes because of the mode frequency factor in Eq.~\ref{eq:response_1} even though the matrix elements $\tau_{x(y,z),\omega_{GS}}$ are nonzero. However, a few peaks appear in the $\tau_{x}-\tau_x$ response below the particle-hole continuum. These are identified as Higgs-like modes because they produce poles in the amplitude-amplitude response functions and correspond to the Higgs-like modes identified in studies of superconductors. The property that they appear below the particle-hole continuum is the key difference from superconductors with short-range interaction.

\begin{figure*}[htbp]
	\includegraphics[width=2.0\columnwidth]{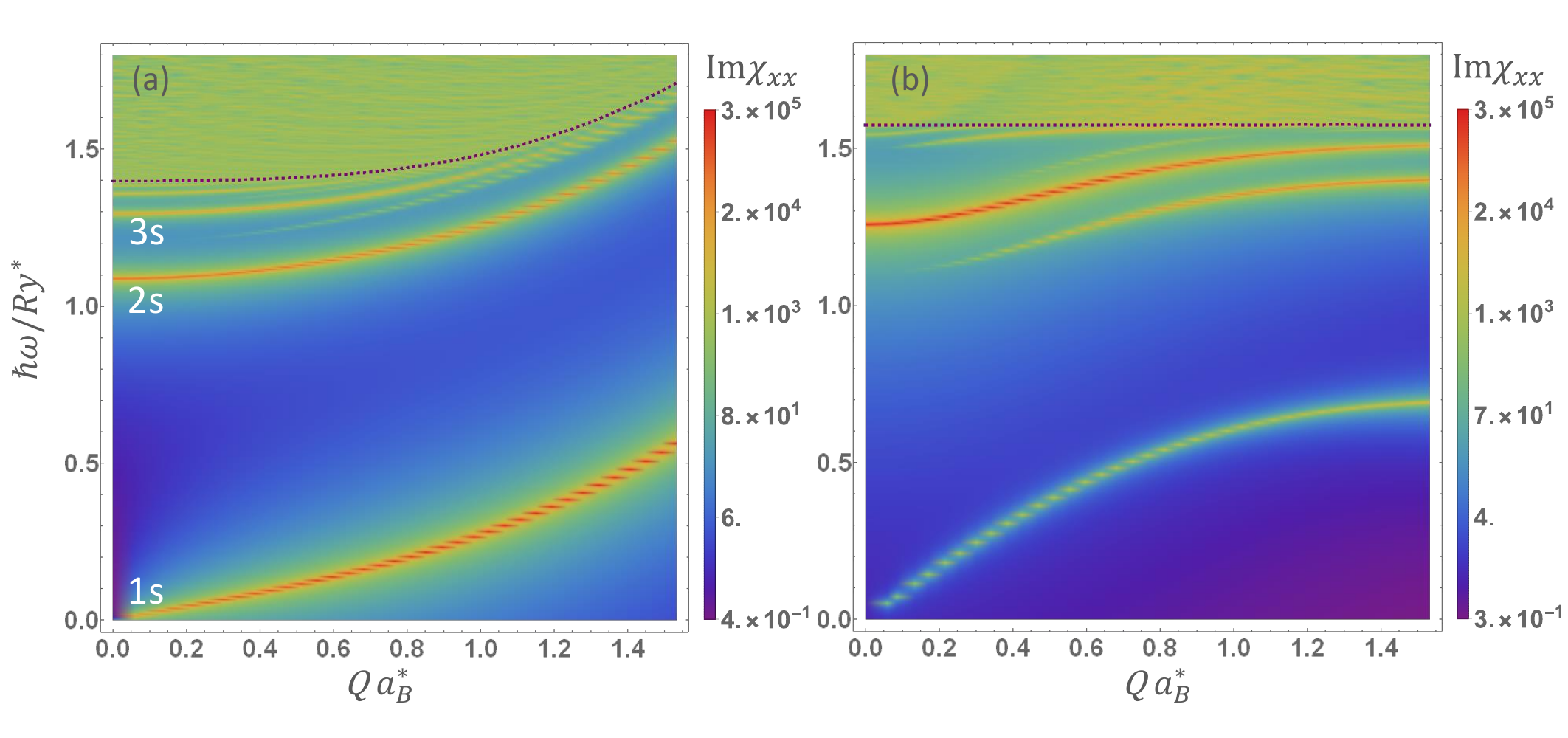}
	\caption{(Color online) Intensity $\text{Im}\chi_{xx}$ as a function of center-of-mass momentum Q and excitation energy $\omega$ in both (a) BEC and (b) BCS regimes.
		The purple dashed line represents the electron-hole continuum. In the BEC regime in (a), three collective mode branches below the continuum dominating the responses are Goldstone mode ($1s$) and Higgs-like modes ($2s$ and $3s$) plotted in Figs.~\ref{Fig:BECwfn}(a) to (c). In the BCS regime in (b), only one gapped Higgs-like mode dominates the response at small $Q$ ($Qa_B^*<0.3$) and the second Higgs-like mode which has lower energy than the first Higgs-like mode, appears as $Q$ increases. 
	}
	\label{Fig:colorplot}	
\end{figure*}

To understand the character of the Higgs-like modes more fully,
we examine the low carrier density limit in which $u_{\vec{k}}$ has small values at all $\vec{k}$.  
From Eq.~\ref{eq:tau2} it follows that, to lowest order in $u_{\vec{k}}$,
\begin{equation}
\begin{split}
&\mathcal{K}^{(\pm)}=\delta_{\vec{k},\vec{p}}(E_{\vec{k}}+E_{\vec{k}+\vec{Q}})-
\frac{1}{A}U(\vec{k}-\vec{p}),\\
&\tau_{x,y}(\vec{Q})=\sum_{\vec{k}}x_{\vec{k}}(\vec{Q}),\\
&\tau_{z}(\vec{Q})=\sum_{\vec{k}}(u_{\vec{k}}+u_{\vec{k}+\vec{Q}})x_{\vec{k}}(\vec{Q}).
\end{split}
\end{equation} 
In the dilute limit the matrices $\mathcal{K}^{(+)}= \mathcal{K}^{(-)}$ reduce
to the two-particle electron-hole relative motion
Hamiltonian matrices at center-of-mass wavevector $\vec{Q}$. 
This very dilute exciton condensate limit becomes a standard two-dimensional hydrogen-like problem where each excitation can be characterized by atomic-like orbitals, such as $1s, 2s, 2p$, etc. 
Note that in this limit, the $\chi_{zz}$ response weighting factor projects out
relative-motion states that are orthogonal to the pair state that is macroscopically occupied in the ground state - 
$1s$ hydrogenic pair states in the Coulomb interaction case. 
We have computed the momentum space wavefunctions, i.e., eigenvectors of $\mathcal{K}^{(+)}\mathcal{K}^{(-)}$, for the six lowest-energy collective modes in the BEC regime at $\vec{Q}=0$ (Fig.~\ref{Fig:Spectra}(a)) and find that they resemble hydrogenic atomic orbitals very well, as shown in Fig.~\ref{Fig:BECwfn}.
The energy sequence of these modes is $1s, 2p, 2s, 3d, 3p$, and $3s$ where $1s$ is the gapless Goldstone mode and $2s$ and $3s$ are Higgs-like modes with peaks in $\chi_{xx}$ responses (denoted with arrows in Fig.~\ref{Fig:Spectra}(a)). Note that $2p$, $3p$, and $3d$ states are doubly degenerate and only one state of each doublet is shown in Figs.~\ref{Fig:BECwfn}(d)--(f).
In this way we have found that the large-weight amplitude 
response corresponds to the addition of an electron-hole pair to the system, not 
in the $1s$ pair state which is condensed, but in higher-energy orbitals.
The lowest-energy high-weight state in the BEC limit at $Q=0$ corresponds to adding an electron-hole pair in a $2s$ state,
which in two dimensions has a binding energy relative to the particle-hole continuum
that is smaller by a factor of $9$.  The second-highest weight state corresponds to the 
$3s$ state and higher $n$ excitations are not fully identifiable only because 
of the finite density of the momentum space grids used in our calculations.
In a SIXC, therefore, the gapped Higgs-like modes are excitations in which one electron-hole pair is 
added in a state that is orthogonal to the pair state present in the condensate in the BEC regime.   
As we see in Fig.~\ref{Fig:Spectra} (c), at finite $\vec{Q}$, the Goldstone mode also makes a nonzero contribution 
to the pair amplitude response which is even larger in magnitude due to the mixing of phase and amplitude fluctuations. 
The $2s$-like Higgs-like mode still has a very large weight in the
spectra and is located below the electron-hole continuum even in the 
large wave vector $\vec{Q}a_B^*=1$ case.  
In the low exciton density limit, the wavevector dependence of the 
Goldstone collective mode is consistent with the Bogoliubov theory of weakly interacting bosons.
Figure~\ref{Fig:colorplot}(a) shows the intensity of $\text{Im}\chi_{xx}$ as a function of $Q$ and excitation energy $\omega$ in the BEC regime.
We can identify three dominant branches of collective modes having large weight in amplitude-pair fluctuations. Note that $3s$ becomes fainted at large Q due to its closeness to the continuum (purple dashed line). The $2s$-like Higgs-like mode is one of our main findings in the SIXC.

As the BCS regime at large exciton densities is approached, 
the spectra of the Higgs-like modes change. 
In Figs~\ref{Fig:Spectra}(b) and (d), we find that different collective modes have
a larger weight in the response function as
wave vector increases. At zero wave vector, we identify the first Higgs-like mode shown in Fig.~\ref{Fig:Spectra}(b) and 
find that its qualitative interpretation as the excitation of a noncondensed pair is unchanged.  As
$\vec{Q}$ varies, a second peak belonging to a different excitation energy below the first Higgs-like mode appears, 
and the new Higgs-like mode shows higher peak at $\vec{Q}$ increases, as illustrated in Fig~\ref{Fig:Spectra}(d). 
Figure~\ref{Fig:colorplot}(b) shows that only one prominent gapped Higgs-like mode below the continuum has large intensity in $\text{Im}\chi_{xx}$ response functions
at small wavevector besides the Goldstone mode. As $Q$ increases, the second gapped Higgs-like mode below the first Higgs-like mode appears which is qualitatively different from the BEC case. 
We suspect that the large weight in the first Higgs-like mode spreads to other modes as the first Higgs-like mode disperses closer to the flat particle-hole continuum, while the hydrogenic $2s$-like Higgs-like mode disperses similar to the continuum in the BEC regime.
We, nevertheless, find that even in the BCS regime
the SIXC supports Higgs-like modes below the particle-hole continuum.
This behavior is in contrast to the case of the BCS models commonly used for superconductors in which 
Higgs-like modes are located exactly at the edge of the particle-hole continuum $2\Delta$ \cite{Varma2015}. 
The Higgs-like modes here are distinct modes, higher in energy than the Goldstone modes but 
still in the excitation gap.  
The source of the difference is the nature of the attractive interaction between electrons and holes, which supports several bound states. The spectrum of amplitude fluctuation can reflect the spectrum of collective particle-hole excitations, including bound states, if any. If we replace the interlayer Coulomb potential by a $\delta$-function attractive interaction with a cutoff, as commonly employed in the theory of superconductivity, the Higgs-like modes evolve into resonances at the bottom of the particle-hole continuum (shown in Appendix.~\ref{App2}) -- the resonances that have \cite{Klein1980,Varma1982,Arovas2011,Barlas2013,Shimano2013,Shimano2014,Sakuto2014,Volovik2014,sherman2015higgs,Varma2015} been identified as the Higgs-like modes of superconductors.  


\section{Discussion}
\label{sec:discussion}

In this paper we have applied time-dependent mean-field theory to spatially indirect exciton condensates
with the goal of identifying collective modes associated with quantum fluctuations in the electron-hole pair amplitude.
We find that in the low exciton density BEC regime the strongest response to Higgs-like perturbations is one 
in which an electron-hole pair is added in a state that is orthogonal to the pair state present in the ground-state condensate.
This interpretation retains qualitative validity when the exciton density is increased and the BCS limit is approached.  
These findings shed new light on previous work that has studied Higgs-like modes in superconductors, 
in which the Higgs-like response appears, mysteriously perhaps, at the edge of the particle-hole continuum.
In light of the present calculations it is clear that this property just reflects the absence in the BCS models used for these 
studies of a higher-energy electron-electron pair bound state and begs the question as to whether or not higher-energy bound states do exist in some superconductors.  Since Higgs-like excitations in superconductors change the 
total electron number, they can be observed only indirectly \cite{Amo2009,Benfatto2016,Measson2018,Shimano2018,Giorgianni2019,Shimano2019,Measson2019,ReviewShimano}. One possible strategy to detect these higher-energy bound states where they are 
suspected is to look for resonant features in the bias voltage dependent subgap currents of
Josephson junctions.

Two different cases need to be distinguished when discussing the detection of Higgs-like modes.  When a spatially indirect exciton 
condensate is formed from equilibrium populations of electrons and holes in two separate layers, the operator $\tau_x$
corresponds to tunneling between layers.  The presence of a spatially indirect exciton condensate or incipient condensate 
then appears as an anomaly in the interlayer tunneling current-voltage relationship near zero bias \cite{Eisenstein2004,nandi2012exciton,Kim2017,Dean2017,Tutuc2018,Wang2019}. We anticipate that Higgs-like modes will appear as 
finite-bias voltage anomalies at energies below the particle-hole continuum.

The case in which an exciton condensate is formed in quasiequilibrium systems of electrons and holes, 
either in the same layer or in adjacent layers, generated by optical pumping is perhaps simpler experimentally.
In this case the coherent excitons are routinely \cite{OpticalShimano} examined by measuring the photoluminescence (PL) signal.
Emitted photons with energy $\hbar \omega$ can be generated by transitions between initial $N$-exciton states and 
final $N-1$ exciton states which satisfy
\begin{widetext}
\begin{equation} 
\hbar \omega = E_{i}(N)-E_{j}(N-1) = \mu_{ex} + [E_{i}(N)-E_0(N)]-[E_{j}(N-1)-E_{0}(N-1)].
\label{Eq:PL}
\end{equation}
\end{widetext}
The matrix elements for these processes are proportional to the operator $\tau_{x}$, which changes the 
number of electron-hole pairs present in the system by one, and can therefore generate Higgs-like excitations.  
In Eq.~\ref{Eq:PL}, $\mu_{ex}$ is the chemical potential of excitons which is non-zero in non-equilibrium
condensed exciton systems, $E_{i0}(N) = E_{i}(N)-E_0(N)$ is the excitation relative to the ground state in the 
$N$-exciton initial state, and $E_{j0}(N-1)= E_{j}(N-1)-E_{0}(N-1)$ is the excitation energy relative to the ground state 
in the $N-1$ exciton final state.  A similar analysis applies in the case of polariton condensates in which the 
exciton system is coupled to two-dimensional cavity photons \cite{Littlewood2011}.
The PL spectrum consists of a segment for which $\hbar \omega > \mu_{ex}$ due to thermal excitations in the initial state and a 
so-called ghost segment in which $\hbar \omega < \mu_{ex}$ due to excitations being generated in the final state when 
the exciton number changes.  Because they have a high energy, Higgs-like modes are not likely to be thermally populated,
but they can be visible in the ghost mode spectrum when exciton-exciton interactions are strong.
Indeed very recent work \cite{Snoke2019} which appeared as this paper was under preparation 
has claimed that a Higgs-like excitation is present in the PL spectrum of a polariton condensate
at energy $\hbar \omega = \mu_{ex} - E_{Higgs}$, and has made the numerical observation that $E_{Higgs}$ is 
close to the energy difference between the cavity-dressed $2s$ and $1s$ excitonic bound states. 
The present paper appears to explain this observation.

\section{Acknowledgments}

F.X. thanks M. Lu and X.-X. Zhang for helpful discussions.
This work was supported by the Army Research Office
under Award No. W911NF-17-1-0312 (MURI) and by the Welch
Foundation under Grant No. F1473.
F.X. acknowledges support under the Cooperative Research Agreement between the University of Maryland and the National Institute of Standards and Technology Physical Measurement Laboratory, Award No. 70NANB14H209, through the University of Maryland.
F.W. is supported by the  Laboratory for
Physical Sciences.

\appendix

\section{Case of short-range interlayer interaction}
	\label{App2}
	\begin{figure}[htbp]
		\includegraphics[width=1.0\columnwidth]{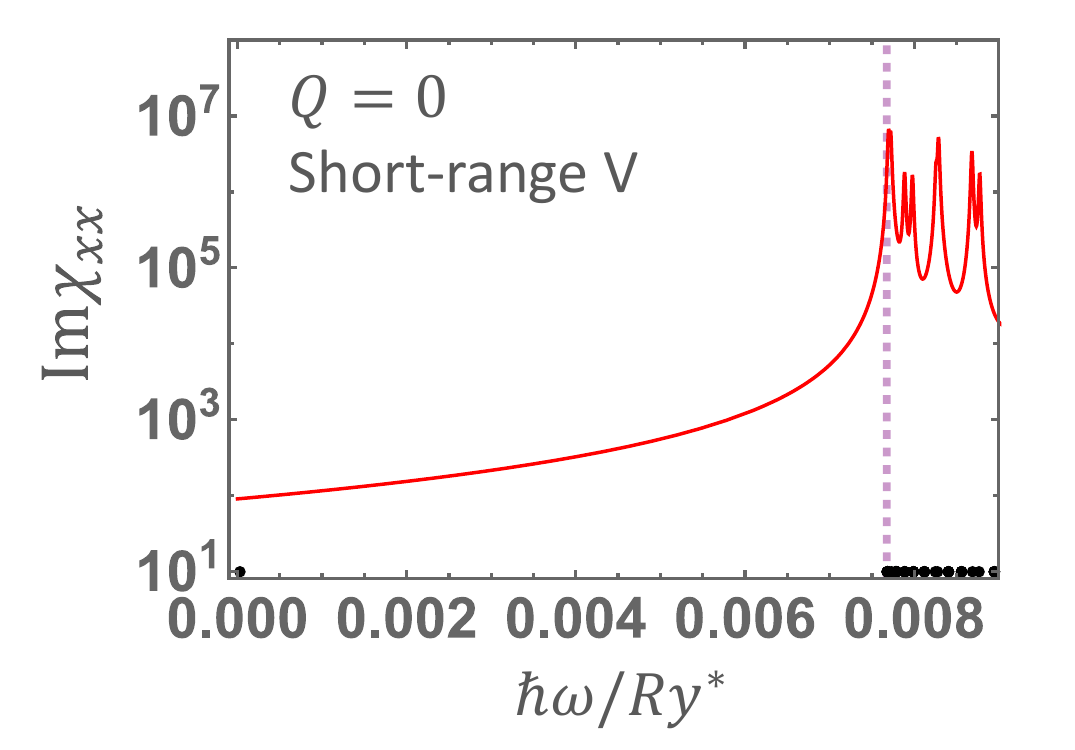}
		\caption{(Color online) 
			Spectra of the magnitude of the imaginary part of the $\tau_{x}-\tau_{x}$ response functions at $Q=0$ with Dirac-$\delta$ short-range interaction. 
			Black dots along the $x$ axis represent the positive collective excitation energies which are square roots of eigenvalues of $\mathbf{\Gamma}$ (Eq.~\ref{eq:gamma}) and $\mathcal{K}^{(+)}\mathcal{K}^{(-)}$ (Eq.~\ref{eq:eom}).
			The purple dashed line denotes the location of electron-hole continuum, i.e., the minimum of $E_{\vec{k}}+E_{\vec{k}+\vec{Q}}$. 
		}
		\label{Fig:deltaSpectra}
	\end{figure}
	By replacing the long-range Coulomb interaction between electrons and holes $U(\vec{q})\propto e^{-qd}/q$ with a Dirac-$\delta$ short-range interaction $U(\vec{q})\propto\delta(\vec{q})$, we find that only the gapless Goldstone mode exists below the continuum and Higgs-like modes are right at the edge of the electron-hole continuum shown in Fig.~\ref{Fig:deltaSpectra}. This is very similar to the case of a BCS superconductor where the Higgs-like mode is located exactly at the particle-hole continuum. Note that the mean-field gap $\Delta_{\vec{k}}$ is a constant $2\Delta_0$ and the continuum is $2\Delta_0$ at $Q=0$. 
\section{explicit expressions for $\mathcal{E}_{\vec{k},\vec{p}}(\vec{Q})$ and $\Gamma_{\vec{k},\vec{p}}(\vec{Q})$}
\label{App1}
Below are explicit expressions for $\mathcal{E}_{\vec{k},\vec{p}}(\vec{Q})$ and $\Gamma_{\vec{k},\vec{p}}(\vec{Q})$, which appear in the energy variation $\delta E^{(2)}$ in Eq.~(\ref{eq:EF}).
\begin{widetext}
	\begin{equation}
	\begin{split}
	\mathcal{E}_{\vec{k},\vec{p}}(\vec{Q})&=\delta_{\vec{k},\vec{p}}(\zeta_{\vec{k}+\vec{Q}}-\zeta_{\vec{k}}+E_{\vec{k}}+E_{\vec{k}+\vec{Q}})\\
	&+\frac{1}{A}\big[V(\vec{Q})-V(\vec{k}-\vec{p})\big]
	(u_{\vec{k}}u_{\vec{p}}v_{\vec{k}+\vec{Q}}v_{\vec{p}+\vec{Q}}
	+v_{\vec{k}}v_{\vec{p}}u_{\vec{k}+\vec{Q}}u_{\vec{p}+\vec{Q}})\\
	&-\frac{1}{A}U(\vec{Q})(v_{\vec{k}}u_{\vec{p}}u_{\vec{k}+\vec{Q}}v_{\vec{p}+\vec{Q}}+u_{\vec{k}}v_{\vec{p}}v_{\vec{k}+\vec{Q}}u_{\vec{p}+\vec{Q}})\\
	&-\frac{1}{A}U(\vec{k}-\vec{p})(u_{\vec{k}}u_{\vec{p}}u_{\vec{k}+\vec{Q}}u_{\vec{p}+\vec{Q}}+v_{\vec{k}}v_{\vec{p}}v_{\vec{k}+\vec{Q}}v_{\vec{p}+\vec{Q}}),\\
	\Gamma_{\vec{k},\vec{p}}(\vec{Q})&=\frac{1}{A}\big[V(\vec{Q})-V(\vec{k}+\vec{Q}-\vec{p})\big]
	(u_{\vec{k}}u_{\vec{p}}v_{\vec{k}+\vec{Q}}v_{\vec{p}-\vec{Q}}
	+v_{\vec{k}}v_{\vec{p}}u_{\vec{k}+\vec{Q}}u_{\vec{p}-\vec{Q}})\\
	&-\frac{1}{A}U(\vec{Q})(v_{\vec{k}}u_{\vec{p}}u_{\vec{k}+\vec{Q}}v_{\vec{p}-\vec{Q}}
	+u_{\vec{k}}v_{\vec{p}}v_{\vec{k}+\vec{Q}}u_{\vec{p}-\vec{Q}})\\
	&+\frac{1}{A}U(\vec{k}+\vec{Q}-\vec{p})
	(v_{\vec{k}}u_{\vec{p}}v_{\vec{k}+\vec{Q}}u_{\vec{p}-\vec{Q}}
	+u_{\vec{k}}v_{\vec{p}}u_{\vec{k}+\vec{Q}}v_{\vec{p}-\vec{Q}}),
	\end{split}
	\end{equation}
\end{widetext}	
where $u_{\vec{k}}$ and $v_{\vec{k}}$ are defined in Eq.~(\ref{eq:ukvk}), and $V(\vec{Q})$ and $U(\vec{Q})$ are respectively intralayer and interlayer Coulomb interactions.

\bibliography{Higgs}{}
\bibliographystyle{apsrev4}
\end{document}